\documentclass[aps,pre,twocolumn,superscriptaddress]{revtex4-1}
\usepackage[english]{babel}
\usepackage{graphicx}
\usepackage{amsmath}
\usepackage{amsfonts}
\usepackage[usenames,dvipsnames]{color}

\usepackage{hyperref}
\usepackage{rotating}
\usepackage{booktabs} 
\usepackage{slashed}
\usepackage{transparent}
\usepackage{color}
\usepackage{soul}
\usepackage{tikz}
\usepackage{balance}

\usetikzlibrary{external}
\usetikzlibrary{arrows}

\usetikzlibrary{patterns}
\usetikzlibrary{shapes,snakes}

\newcommand{\dd}{\; \mathrm{d}}

\DeclareMathAlphabet{\mathbb}{U}{bbold}{m}{n}

\newcommand{\balancecolumns}{%
	\vfill\eject
	\global\@colht = \textheight
	\global\ht\@cclv = \textheight}

\newcommand{\bs}[1]{\boldsymbol{#1}}

\usepackage[normalem]{ulem}

\usepackage{hyperref}
\usepackage{nameref}

\newcommand{\probability}{\mathbb{P}}
\newcommand{\expectation}{\mathbb{E}}

\DeclareMathOperator{\arccot}{arccot}
\DeclareMathOperator{\arccsc}{arccsc}

\makeatletter
\def\maketitle{
	\@author@finish
	\title@column\titleblock@produce
	\suppressfloats[t]}
\makeatother

\begin{document}

\title{A universally applicable approach to connectivity percolation
}

\author{Fabian Coupette}
\email{f.coupette@leeds.ac.uk}
\affiliation{Institute of Physics, University of Freiburg, Hermann-Herder-Stra{\ss}e 3, 79104 Freiburg, Germany}
\affiliation{School of Mathematics, University of Leeds, Leeds LS2 9JT, United Kingdom}
%\affiliation{email: f.coupette@leeds.ac.uk, tanja.schilling@physik.uni-freiburg.de}
\author{Tanja Schilling}
\email{tanja.schilling@physik.uni-freiburg.de}
\affiliation{Institute of Physics, University of Freiburg, Hermann-Herder-Stra{\ss}e 3, 79104 Freiburg, Germany}

\date{\today}

\begin{abstract}
	
	Percolation problems appear in a large variety of different contexts ranging from the design of composite materials to vaccination strategies on community networks. 
	The key observable for many applications is the percolation threshold. Unlike the universal critical exponents, the percolation threshold depends explicitly on the specific system properties. 
	As a consequence, theoretical approaches to the percolation threshold are rare and generally tailored to the specific application.
	
	Yet, any percolating cluster forms a discrete network the emergence of which can be cast as a graph problem and analysed using branching processes.
	We propose a general mapping of any kind of connectivity percolation problem onto a branching process which provides rigorous lower bounds of the percolation threshold. 
	These bounds progressively tighten as we account for more local structure. 
	We showcase our approach for different continuum problems finding accurate predictions with almost no effort. 
	Our approach is based on first principles, does not require fitting parameters, and reproduces all exactly known percolation thresholds. 
	As such it offers an important theoretical reference in a field that is dominated by simulation studies and heuristic fit functions.   
\end{abstract}

\maketitle
Percolation describes the formation of giant components in complex systems \cite{stauffer1991introduction,grimmett1999percolation}. 
Originally proposed to describe water permeating a rock through the emergence of a system spanning cavity network \cite{broadbent1957percolation,hammersley1957percolation,sykes1964exact,kirkpatrick1973percolation}, percolation theory has been applied in a broad variety of different contexts \cite{sahimi1994applications,saberi2015recent} such as the design of composite materials \cite{flandin1999anomalous,sandler2003ultra,kyrylyuk2008continuum,bauhofer2009review,otten2009continuum,nan2010physical,chatterjee2010connectedness,nigro2013quasiuniversal,coupette2021nearest,coupette2021percolation,chatterjee2022geometric}, the analysis of complex networks \cite{moore2000epidemics,cantwell2019message,allard2019percolation,li2021percolation,newman2023message,bianconi2024theory}, and transport through porous media \cite{rossen1990percolation,berkowitz1993percolation,torquato2002random,hunt2014percolation,hunt2017flow}. 
The phenomenon attracted particular attention due to its resemblance of a thermodynamic phase transition \cite{coniglio1981thermal,aizenman1987sharpness,georgii1996phase,jacobsen2014high,duminil2019sharp}, where the percolation probability acts as an order parameter and the mean cluster size can be interpreted as a susceptibility with characteristic power-law exponents describing their scaling behavior in the vicinity of the critical point, i.e., the percolation threshold. 
While those critical exponents coincide for all standard percolation problems set in the same spatial dimension \cite{stauffer1979scaling,essam1980percolation,vicsek1981monte,lee1990monte,smirnov2001critical,gracey2015four}, the percolation threshold itself sensitively depends on the intricacies of the system. 
As a consequence, theoretical approaches are mostly tailored to specific applications \cite{coniglio1977pair,kesten1980critical,desimone1986theory,cardy2003exact,kyrylyuk2008continuum,wu2010critical,chatterjee2010connectedness,karrer2010message,widder2021generating} and straightforward simulation is the primary tool of choice for the accurate determination of critical parameters \cite{rintoul1997precise,lorenz1998precise,lorenz2001precise,miller2003competition,mertens2012continuum,jacobsen2014high,schilling2015percolation,xu2021critical}.
Yet, the predictive power of these approaches is limited.
\newline
\newline
Connectivity percolation has been studied with a multitude of different prefixes such as lattice, continuum, directed, dynamic, protected, explosive or bootstrap \cite{dorogovtsev2008critical,araujo2014recent,saberi2015recent,lee2018recent}.
In spite of this variety of flavours, percolation inherently is always a graph problem. 
Even if particles move continuously in space, the connectivity relationship between the particles in the system can be expressed as a graph with each vertex representing a particle and edges corresponding to a connection between the particles represented by the incident vertices. 
Thus, each configuration $\gamma$ of the system translates into a network which may be independently analysed for the existence of a giant component. 
Therefore, all percolation problems have a unified foundation which we will exploit in the following in order to develop a controlled approximation for the percolation threshold irrespective of the particular flavour of percolation problem.

Within this article, we generalize concepts known from discrete network percolation to any kind of connectivity percolation problem. This allows us to construct a hierarchy of rigorous lower bounds to the percolation threshold for continuum percolation problems which are substantially tighter than what conventional approaches such as connectedness percolation theory can provide. Most notably however, the same criterion that generates these lower bounds also generates the exact solutions to all exactly solved discrete percolation problems \cite{coupette2022exactly}. Thus, we present a first-principle universally applicable method to estimate connectivity percolation thresholds. As percolation thresholds play an important role across fields and scales from material design to epidemic outbreaks, a reliable protocol to estimate critical transitions with a robust theoretical foundation is an indispensable tool. Branching processes form the key element in our construction.

\section{Percolation as a branching process}
For illustrative purposes consider an infinite connected graph $G$. We define a percolation problem on $G$ by assigning a degree of freedom to each edge, i.e., each edge is either open or closed. Two vertices are considered connected if there is a path of open edges linking them. We assign an arbitrary vertex of the graph as the origin, $\mathcal{O}$, and
call a vertex open if it is connected to the origin.
The percolation probability $\Theta$ is defined as the probability that the origin is part of an infinite cluster of  connected vertices \cite{grimmett1999percolation}. 
If the model is taking parameters $\bs{p} \in \Lambda$ from a parameter space $\Lambda$ (i.e.~if the probability of an edge being open depends on the values of the parameters $\bs{p}$) then one can ask at which values an infinite cluster forms.  The ``percolation threshold'' (or more generally, the critical manifold) is defined as the boundary of the set
\begin{align}
	\{ \bs{p} \in \Lambda : \Theta(\bs{p}) = 0 \} \;.
	\label{eq:critical_manifold}
\end{align}
This a natural generalization of $\sup\{p :
\Theta(p) = 0\}$ used for standard lattice percolation problems
\cite{grimmett1999percolation} to higher-dimensional parameters
spaces.

To introduce a method to predict percolation thresholds, we now
partition the graph into nested layers, which grow outwards from the origin. For this purpose, we define the $k$-walk neighbourhood $\mathcal{N}_k$ of $\mathcal{O}$ as the sub-graph spanned by all random walks of length $k$ starting at the origin We drop the prefix ``walk'' in the following but note that our definition of $\mathcal{N}_k$ excludes edges between two vertices at a distance $k$ from the origin in contrast to the standard definition of a neighbourhood. 
The graph can be partitioned into mutually disjoint vertex sets, if we group them by the length of the shortest path to the origin.  We use the notation $V(\mathcal{N}_{k})$ for the set of vertices of the $k$-neighbourhood and $V_k = V(\mathcal{N}_k) \setminus V(\mathcal{N}_{k-1})$ for those vertices, that are connected to the origin by a shortest path of length exactly $k$. By construction, a percolating cluster around the origin requires at least one open vertex in each vertex set $V_k$. 
Given a configuration of the system, we define $X_k$ as the number of open vertices in $V_k$ on the sub-graph $\mathcal{N}_k$. 
Accordingly, $X_1$ is the number of neighbours directly connected to the origin, i.e., the number of vertices each connected to the origin through a single open edge. 
The direct neighbours of any of these $X_1$ vertices, apart from the origin, comprise $X_2$.
Continuing to count the direct neighbours, but excluding all vertices that have been visited in previous layers, we construct a sequence $(X_k)_{k \in \mathbb{N}}$ that we call "surface activity sequence". We use the term "activity", because
only open vertices in $V_k$ have the capacity to induce open vertices in $V_{k+1}$. Thus, $X_k$ counts the number of growth sites which quantifies how likely the $k$-th layer is to produce connections to the $k+1$-th layer, a property which we call the "activity" of the layer. 
Consequently, the sequence terminates once $X_{k} = 0$, i.e., there is  a finite connected cluster around the origin. 
The question is now, how the sequence of $X_k$ is related to the percolation threshold. To answer this question, we interpret the problem as a branching process.
\newline
If the system is treelike, one can formally write $(X_k)_{k \in \mathbb{N}}$ as a branching process with integer random variables $\xi_i^k$ describing the offspring of open vertex $i$ in layer $k$
\begin{align}
	X_{k+1} = \sum^{X_{k}}_{i=1} \xi_i^k \quad \text{with} \quad X_0 = 1 \; .
	\label{eq:bp}
\end{align}
There is no dependence between the $\xi_i^k$, because in a treelike system the ``offspring'' (the number of open vertices in $V_{k+1}$) induced by each individual open vertex in $V_k$ are independent random variables. 
This is no longer the case if the graph has loops, because multiple vertices in $V_k$ may be connected to the same vertex in $V_{k+1}$. Yet, we may still cast the problem as a generalized branching process of the form
\begin{align}
	X_{k+1} = \sum^{X_{k}}_{i=1} \xi_i^k(\{\xi^k_j\}_{j<i}) \quad \text{with} \quad X_0 = 1 \; ,
	\label{eq:main}
\end{align}
with $\xi_i^k(\{\xi^k_j\}_{j<i})$ being an integer random variable drawn from the conditional distribution 
of next-level neighbours generated by the $i^{\mathrm{th}}$ open vertex in the $k$-neighbourhood. The key difference to the treelike scenario is that we specify the random variables in an order (i.e.~we label the open vertices in $V_k$ by the numbers $1 \ldots X_{k}$ and sum them in that order) and the results impact the distribution of following random variables within one layer $k$. 
\begin{figure}[t!]
	\includegraphics[width = 0.21\textheight]{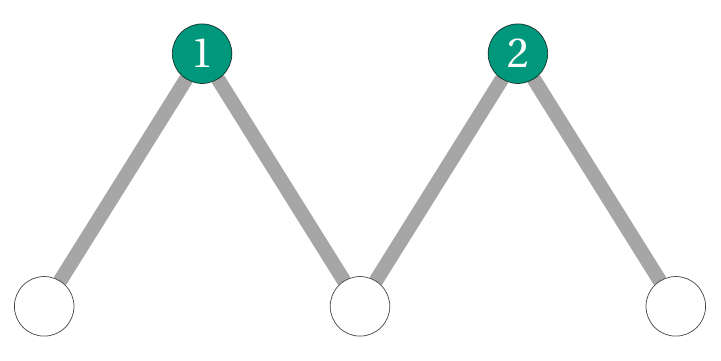}
	\includegraphics[width = 0.036\textheight]{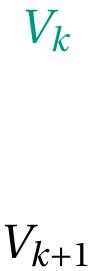}
	\caption{Impact of loops in the transition between layers exemplified for bond percolation with edge probability $p$. Filled circles represent open vertices in $V_k$, open circles are potential branching partners.}
	\label{fig:loops}
\end{figure}
\newline
To illustrate this, consider the bond percolation problem with edge probability $p$ depicted in Fig. ~\ref{fig:loops}: \newline
Two open vertices in $V_k$ can potentially connect to the same vertex in $V_{k+1}$. This induces correlations between $\xi^k_1$ and $\xi^k_2$. The initial random variable depends only on its forward coordination number $\xi^k_1 \sim B(2,p)$ (binomial distribution). However,  
\begin{align*}
	\xi^k_2(\xi^k_1) \sim \begin{cases}
		B(2,p) & \text{if} \; \xi^k_1 = 0 \\
		\frac{1}{2}\left[ B(1,p)+ B(2,p) \right] & \text{if} \; \xi^k_1 = 1 \\
		B(1,p) & \text{if} \; \xi^k_1 = 2 \\
	\end{cases}
\end{align*}, \newline to account for double-counting of the shared vertex. Clearly, $\expectation[\xi^k_{1}+\xi^k_{2}(\xi^k_{1})] = 4 p - p^2$. As any realization of $\xi_1^k$ except for $\xi_1^k = 0$ reduces the expectation of $\xi_2^k$,  the two random variables are necessarily anti-correlated. 
\newline
The order we impose on the specification of random variables can be understood as a breadth first search. 
As a consequence, any vertex in $V_{k+1}$ can only have one parent in $V_k$ so that the graph generated by the modified search algorithm is again treelike. 
Hence, we effectively eliminate the loops at the expense of correlations between the $\{\xi^k_j\}_{1 \leq j \leq X_k}$. The extinction probability $Q$ of the branching process is connected to the percolation probability via $Q = 1 - \Theta$. Thus, we can link the distributions of the random variables $\xi^k_j$ to the percolation threshold.
\newline

At this stage, it is irrelevant whether a particular sequence $(X_k)_{k \in \mathbb{N}}$ was generated by bond percolation on the square lattice, cavities in a porous medium, or carbon nano-tubes dispersed in a polymer matrix.  The percolation threshold depends exclusively on the set of random variables defining the branching process.
We can carry out this mapping for any connectivity percolation problem given a notion of a $k$-neighbourhood. 
Yet, we have only transferred the original complexity onto the definition of the random variables $\xi_i^k$. So what is all this good for?
\newline
\newline
There are two different aspects contributing to the complexity of $\xi_i^k$ on the graph level: vertex degree correlations and loop structure. 
The former can express themselves as, for instance, the friendship paradox \cite{eom2014generalized,cantwell2021friendship} in social networks or the structure of a hard sphere fluid \cite{hansen1990theory}.
Vertex degree correlations can be both, positive as well as negative, and they tend to decay with the distance between layers of the construction. 
In simplified terms, vertex degree correlations account for the $k$-dependence of $\xi^k_i$. 
Loops, on the other hand, exclusively induce negative correlations in the sense which is demonstrated in the caption of Fig.~\ref{fig:loops}, i.e.,
\begin{align}
	\expectation[\xi^k_i \xi^k_{j<i}] \leq \expectation[\xi^k_i]\expectation[\xi^k_{j<i}] \; .
	\label{eq:neg_cor}
\end{align} 
If there are loops, then multiple paths can activate the same surface vertex. This leads to a reduced average surface activity compared to the treelike case \cite{hammersley1957percolation}. 
We are going to exploit this one-sided correlation in the following when constructing a lower bound for the percolation threshold. 

\section{Treelike Systems}

\begin{figure}[t]
	\includegraphics[width = 0.49\textwidth]{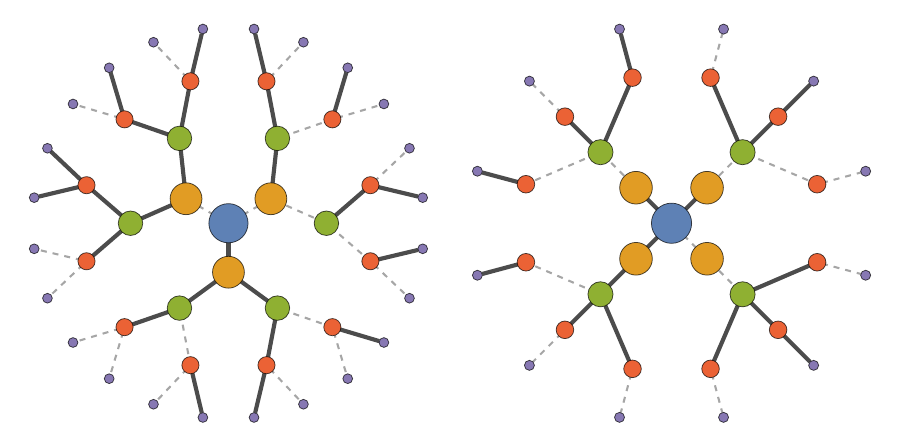}
	\caption{
		Left: Percolation on a regular Bethe lattice -- $p_c = \frac{1}{\langle z \rangle - 1} = \frac{1}{2}$. 
		Right: The impact of vertex degree correlations. Tree with the same mean degree as the left lattice but alternating vertex degree -- $p_c = \frac{1}{\sqrt{3}}  > \frac{1}{2}$. \newline
		Colors and vertex sizes distinguish the vertex sets $V_k$.
		Edges illustrate a realization of a bond percolation model: thick bonds are open, dashed bonds are closed. }
	\label{fig:Bethe_lattices}
\end{figure}

Treelike networks are an important special case because in this case equality holds in eq.~(\ref{eq:neg_cor}). This allows for exact calculation of the percolation thresholds \cite{dorogovtsev2008critical,hamilton2014tight,karrer2014percolation,coupette2022exactly}.
If the percolation problem is defined on a treelike graph the formation of treelike clusters is evident, but there are less obvious instances of treelike network formation such as Erdős–Rényi random graphs in the thermodynamic limit, slender rods in three dimensions, or Poisson processes in infinite dimensions.
Without loops, the probability distributions corresponding to the individual $\xi^{k}_i$ only depend on the distribution of vertex degrees across the system. 
If, furthermore, the network is asymptotically homogeneous, i.e., $\xi^{k}$ converges in distribution to a common random variable $\xi$, 
the problem simplifies to a Galton-Watson branching process \cite{harris1963theory}. 
As a consequence, the percolation problem on such treelike network is critical if 
\begin{align}
	\expectation[\xi] = 1 \; ,
	\label{eq:2nd}
\end{align}
i.e., an active vertex induces on average another active vertex on the next level. The random variable $\xi$ captures the recurring transition from one layer of the construction to the next. For example, percolation on the regular Bethe lattice with coordination number $z = 3$ depicted in Fig.~\ref{fig:Bethe_lattices} (left panel) is generated by $\xi^1_{1} \sim B(z,p)$ (binomial distribution) and $\xi^{k>1}_{i} \equiv \xi \sim B(z-1,p)$ resulting in a critical edge probability $p_c = \frac{1}{z - 1} = \frac{1}{2}$. The initial branching step differs from the rest due to the extra branching partner of the origin, but the percolation threshold depends exclusively on the asymptotic random variable $\xi$.
We previously demonstrated that all exactly solved percolation problems, even bond percolation on the square lattice, can be mapped on a branching process and the critical parameters are always given through a version of eq.~(\ref{eq:2nd}) \cite{coupette2022exactly}. 
If the network features degree correlations the definition of $\xi$ may require an additional coarse-graining step but the criticality condition remains valid. With alternating vertex degrees as shown in the right panel of Fig.~\ref{fig:Bethe_lattices}, we only need to coarse-grain two subsequent branching steps into one to recover a homogeneous branching process with $\xi^{k>1}_{i} \equiv \xi \sim B(1,p) B(3,p)$ and $p_c = \frac{1}{\sqrt{3}}$. This result may equivalently be obtained
as the inverse of the largest eigenvalue $\lambda_1$ of the non-backtracking (Hashimoto) matrix $H$ of the respective lattice 	\cite{hamilton2014tight,karrer2014percolation}. 
$H$ generates non-backtracking walks in the sense that $H^L_{\alpha \beta}$ counts the number of non-backtracking walks comprising $L+1$ edges starting on the directed edge $\alpha$ and terminating at directed edge $\beta$. 
On a tree there is at most one non-backtracking walk between any pair of directed edges. Define the end vertex of $\alpha$ as origin $\mathcal{O}$ and the unit vector selecting the $\alpha$-row of $H$ as $e_\alpha^t$.
Then each non-zero entry of the vector $e_{\alpha}^t H^k$ corresponds to a non-backtracking walk from $\mathcal{O}$ to a vertex within the vertex set $V_k$ of our construction. The probability of each of the walks to be open is $p^k$, so the number of open vertices in $V_k$ (excluded paths reversing $\alpha$) is given by $X_k := ||e_\alpha^t H^k ||_1 p^k$. 
As $k$ goes to infinity, $H^k$ converges to $\lambda_1^k \mathbb{1}$ through von-Mises power iteration (except pathologies). Thus, $(X_k)_{k \in \mathbb{N}}$ diverges for $p \lambda_1 > 1$ and converges to 0 if $p \lambda_1 < 1$ which means $p_c = \lambda_1^{-1}$. If the vertex degrees along a walk are given by a Markov process we apply the same concept to the simpler branching matrix $B_{z z'} : = (z' - 1) \probability(z' | z)$ featuring the probability that a vertex with degree $z$ is adjacent to a vertex with degree $z'$ \cite{goltsev2008percolation}. For the tree in the right panel of Fig.~\ref{fig:Bethe_lattices} we get 
$B=\begin{pmatrix}
	0 & 3 \\  1 & 0 
\end{pmatrix}$ 
and $\lambda_1 = \sqrt{3}$ is readily read off. 
\newline
Another well-studied and exactly solvable example for network percolation is the configuration model, i.e.,  random graphs generated from a given degree sequence. These graphs are generated by randomly matching outgoing edges of different vertices in accordance with their preassigned degree. As a consequence, any vertex is more likely to be connected to a vertex with lots of edges, because there are simply more matching pairs leading to vertices with higher degree. That means, even though the vertex degrees are drawn independently from a distribution $\probability(z)$, the degree-distribution of the neighbour of a randomly chosen vertex is given by the excess degree distribution 
\begin{align}
	\probability'(z) = \frac{z}{\langle z \rangle} \probability(z) \; .
\end{align}
This is the distribution that governs the asymptotic branching process, i.e., the asymptotic random variable $\xi$. Thus, eq.~(\ref{eq:2nd}) yields
\begin{align}
	\expectation[\xi] &=& & p \sum_z z (z-1) \probability'(z) = p \sum_z  \frac{z (z-1)}{\langle z \rangle} \probability(z) \nonumber  \\ &=& & p \frac{\langle z^2  \rangle -\langle z \rangle }{\langle z \rangle} = 1
\end{align}
which is equivalent to the Molloy-Reed criticality criterion \cite{molloy1995critical,coupette2024polydispersity}. In that sense, eq.~(\ref{eq:2nd}) is a generalisation of the Molloy-Reed criterion which can be applied to a much broader class of networks as we show in the following.

In our framework, degree correlations do not fundamentally change the problem because their impact on the asymptotic branching process is readily accounted for. Therefore, we will ignore them for the moment and concentrate on loops, the actual problem.

\section{Networks with Loops}

On a graph with loops, the distribution of $X_{k+1}$ is not entirely defined by $X_k$ but also depends on the position of the active surface vertices relative to each other.
If the length of loops is bounded throughout the system, we can coarse-grain in a way to absorb the largest loop in a single branching step, i.e., we replace the layers $V_k$ by structures sufficiently large to contain the loops, and build the branching process from these structures. Then the solution of the percolation problem is reduced to solving percolation of a decorated Bethe lattice.
That means, without vertex degree correlations, 
\begin{align}
	\expectation[X_k] = 1 \; ,
	\label{eq:s1}
\end{align}
yields the exact percolation threshold if the system features only loops shorter than $2 k+1$.
(The significance of loops of diverging size constitutes the universality class of the percolation problem.)

\begin{figure}[t!]
	\includegraphics[width = 0.21\textheight]{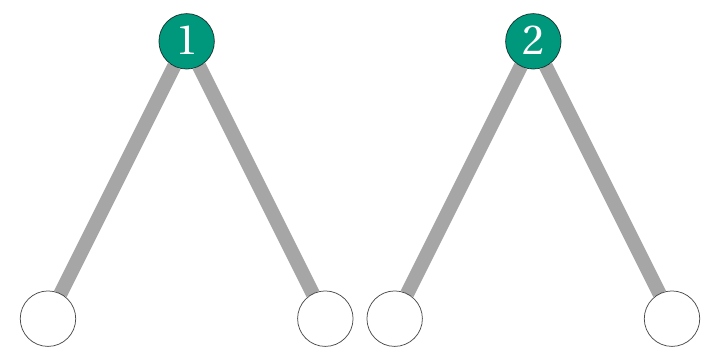}
	\includegraphics[width = 0.036\textheight]{loop_explanation_2}
	\caption{First order treelike reference for the situation of Fig. \ref{fig:loops}. With  $\expectation[X_1] = \expectation[\xi^k_{1}+\xi^k_{2}] = 4 p$, the average size of the next generation is systematically overestimated, yielding a lower bound for the percolation threshold. }
	\label{fig:loops2}
\end{figure}

More importantly, this is also a {\it rigorous lower bound to the percolation threshold} for any system without vertex degree correlations irrespective of its loop structure. The reason for this is the following: a loop of length $2k$ comprises two independent paths of length $k$ which lead to the same vertex. 
If we add the probability of each path to be open (like we would on a tree to compute $\expectation[X_k]$), we count twice the configuration with both paths open simultaneously and hence systematically overestimate the average surface activity. An overestimation of the surface activity implies an underestimation of the percolation threshold.
This is the fundamental observation that our approach is based upon: if we ignore loops, we systematically underestimate the percolation threshold \cite{hammersley1957percolation,hamilton2014tight,karrer2014percolation}.
The bound of eq.~(\ref{eq:s1}) becomes progressively tighter as we increase $k$ giving rise to a hierarchy of approximations. Notice that at order $k$ we also account for loops of length smaller than $2k$ as the effect of those loops is reflected in the paths leading to the leaf nodes. Yet, edges between the vertices in $V_k$ are not included in the walk-neighbourhood $\mathcal{N}_k$.
We call the hierarchy of approximations {\it areal expansion} in analogy to the virial expansion, because in the continuum we will systematically integrate out larger volumes rather than expanding in the number of particles participating in an interaction. 
But why exactly is $\expectation[X_k] = 1$ a rigorous lower bound in any homogeneous system? 
\newline
\newline
We can construct a branching process that generates all configurations of the original system as a subset. For the first order, we outright eliminate all loops, which is to say we ignore all correlations between random variables with the same $k$. The effect is readily illustrated using the example in Fig.~\ref{fig:loops}. The shared vertex is effectively duplicated so that  $\xi^k_1$ and $\xi^k_2$ become independent as shown if Fig.~\ref{fig:loops2}.
Since in the original graph, $\xi^k_1$ and $\xi^k_2$ were anti-correlated, the expectation for the entire generation, i.e., $X_{n+1}$ grows by ignoring those correlations. 
Meanwhile, we can solve the treelike system generated by the simplified branching process exactly -- the criticality condition is eq.~(\ref{eq:2nd}).  Finally, the substitute branching process generates all of the original configurations with the corresponding probability, the difference only being additional configurations that original system did not entail. 
But, the plain addition of opportunities for the cluster containing the origin to grow can only decrease the percolation threshold. This last part is particularly nicely showcased by the example of long range percolation on the integer line discussed in the supplementary material (SM) . 
Putting everything together, we find that in any homogeneous system, a mean non-backtracking coordination number of 1 is a necessary requirement for percolation. 
\newline
\newline
This result is already established in continuum percolation (albeit in rather different terms), because the second virial approximation is known to provide a lower bound to the percolation threshold and its calculation ultimately boils down to eq.~(\ref{eq:s1}) for $k=1$. However, there are two key advantages of the branching process perspective. 
Firstly, we can observe the subtle difference between eq.~(\ref{eq:s1}) and eq.~(\ref{eq:2nd}). 
The percolation threshold depends on the expectation of the asymptotic random variable $\xi$ rather than the average number of neighbours of the origin. The origin is not representative for the asymptotic growth as the origin has no predecessor and thus more directions to branch to. Indeed, this is in many instances the main reason for the quantitative inaccuracy of the second virial approximation.
Secondly, we can systematically improve the approximation by incorporating loops up to a given length while maintaining a rigorous lower bound. This is in crass contrast to the virial approximation which is entirely uncontrolled beyond the second order as is exemplified for a simple Poisson process in the SM.
For $k=2$, we correctly incorporate loops comprising up to four edges. 
As $k$ goes to infinity we eventually integrate out the entire system and the lower bound necessarily becomes arbitrarily tight. Naturally, we cannot compute $X_k$ analytically for large $k$ in complicated systems so that, regarding the exact value of the percolation threshold, we do not gain anything compared to a straightforward simulation. 
The strength of our approach is that we can compute reliable lower bounds with decent precision and little effort. 
Moreover, analytical results for low orders of the construction enable us to directly characterize the impact of model parameters on the percolation threshold. 
\section{Applications}

\subsection{Fully penetrable spheres}
To demonstrate the virtues of our approach, consider one of the most fundamental continuum percolation problems: points randomly distributed in $\mathbb{R}^3$ with a prescribed number density $\rho$. 
Points are connected if their separation is smaller than a threshold distance $d$. Above a critical dimensionless density $\rho_c d^3 \approx 0.6530$ \cite{lorenz2001precise}, the system almost surely contains an infinite cluster of connected points. 
While established simulation techniques allow for an accurate computation of this value, theoretical approaches like connectedness percolation theory which utilizes liquid state theory for connectivity are generally inconclusive. 
The virial series cannot be truncated at any accessible order \cite{alon1990systematic} and liquid state closures to the connectedness Ornstein-Zernike equation generate the wrong diagrammatic expansion \cite{coniglio1977pair,chiew1983percolation,desimone1986theory,chiew1989connectivity,coupette2023percolation}. 
Other methods which provide accurate predictions are either heuristic in nature, based on unjustifiable assumptions or tailored to specific systems \cite{alon1991new,garboczi1995geometrical,coupette2021nearest,coupette2020continuum}. 
With all these approaches, it is extremely hard to \textit{a priori} estimate the accuracy of the result which renders their predictive power virtually inexistent. 
\begin{figure}
	\includegraphics[width = 0.49\textwidth]{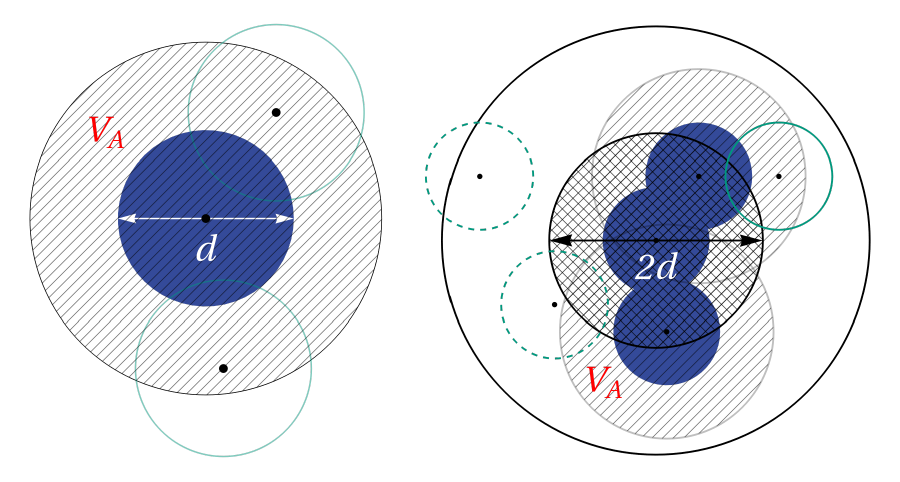}
	\caption{Branching process for fully penetrable disks of diameter $d$. 
		Left: Areal order $k=1$. The blue particle is isolated unless there is a particle centered in the active volume $V_A$ (hatched). 
		Right: Areal order $k=2$. We fix the configuration within a disk of radius $(k-1)d$.  The cluster nucleus remains finite unless there is at least one particle in the new active volume $V_A$ (hatched). 
		The crossed area is already specified so that it does not contribute to $V_A$. 
	}
	\label{fig:continuum_branching}
\end{figure}
\newline
The areal approximation has a straightforward continuum formulation. 
We decompose the probability $p(\bs{r}_1,\bs{r}_2)$ that particles positioned at $\bs{r}_1$ and $\bs{r}_2$ are connected into functions $p_k$, which describe the connection probability with the additional constraint that the shortest path (on the graph level) between the respective particles has length $k$
\begin{align}
	p(\bs{r}_1,\bs{r}_2) = \sum_{k = 1}^{\infty} p_k(\bs{r}_1,\bs{r}_2) \;.
\end{align}
$p_k(\mathcal{O},\bs{r})$ effectively describes the probability density that a particle at $\bs{r}$ is activated in the $k^{\mathrm{th}}$ branching step from the origin. 
Accordingly, the average number of particles activated in the $k^{\mathrm{th}}$ branching step $X_k$ is given by
\begin{align}
	\expectation[X_k] = \int \dd \bs{r} \; \rho(\bs{r}) p_k(\mathcal{O},\bs{r}) \;.
	\label{eq:exk}
\end{align}
Combining eq.~(\ref{eq:s1}) and eq.~(\ref{eq:exk}), we find a sequence $(\rho_c^k)_{k \in \mathbb{N}}$ of rigorous lower bounds for the percolation threshold of a continuum percolation problem. \newline

However, the graph distance is not the natural length scale for a continuum system. 
We can switch to euclidean distance by adapting our notion of a $k$-neighbourhood: instead of referring to the particles that can be reached by a random walk of $k$-steps from the origin on the connectivity graph, we include all particles that can be reached by a random walk that does not exit the ball $\mathcal{B}_{k d}(\mathcal{O})$ (radius $k d$ around the origin).  
The cluster containing the origin is almost surely finite if there is not on average at least one particle outside of that ball that is directly connected to the $k$-neighbourhood of the origin. 
Now we define $\expectation[X_{k+1}]$ as the average number of particles in direct contact with the $k$-neighbourhood outside of the $kd$-ball averaged over all system configurations. 

Applying our approach to the sphere model (generally all non-interacting models) is particularly simple: due to a homogeneous density distribution,
\begin{align}
	\expectation[X_{k+1}] = \rho \, \expectation[V_A(kd)] \; ,
\end{align}
with the active volume $V_A(kd) \subset \mathcal{B}_{kd}(\mathcal{O})^c$ of a configuration defined as the volume outside of the $kd$-ball in which a probe particle would intersect the $k$-neighbourhood of the origin (see Fig.~\ref{fig:continuum_branching}). 
The active volume comprises the set of positions which have the capacity to further extend the cluster that contains the origin. 
For spheres, $V_A(0)$ is the excluded volume of the origin yielding
\begin{align}
	\rho^1_c  = \frac{1}{V_A(0)} = \frac{3}{4 \pi d^3}  \; .
	\label{eq:2ndvir}
\end{align}
This is equivalent to the second virial approximation as required for consistency with exactly solvable models. 
\begin{table}[t!]
	\centering
	\caption{Lower bounds for the percolation threshold calculated with the areal and virial approximations, respectively. The virial calculations are presented in the SM.  	 } 
	\label{tab:sphere_results}
	\begin{ruledtabular}
		\begin{tabular}{c c c c} \toprule
			\bfseries order $k$ &  \bfseries areal $\rho^{kd}_c d^3$ & \bfseries ar. half-space $\tilde{\rho}^{kd}_c d^3$ & \bfseries virial \\\midrule
			1 & 0.2387  & 0.4775 &  --- \\ 
			2 & 0.3468 & 0.5502 & 0.2387 \\
			3 & 0.4375 & 0.5959  & ---\\
			4 & 0.4874 & 0.6099  & 0.3001 \\
			\hline
			Simulation & 0.6530 & 0.6530 & 0.6530 \\\bottomrule
		\end{tabular}
	\end{ruledtabular}
\end{table} 
Yet, the third virial order accounts only for the addition of triangles which does not yield a real percolation threshold (see SM).
Conversely, the second areal approximation accounts for all configurations within the ball $\mathcal{B}_{d}(\mathcal{O})$ which may contain arbitrarily many pairwise interacting particles. Thus, the areal expansion is not a density expansion but rather expands in the length of incorporated loops. That generally makes it much harder to make analytic progress given strongly interacting systems, but for simple Poisson processes, the areal framework is a natural and intuitive way to decompose configuration space.
%Moreover, the areal approximation by design provides lower bounds of the percolation thresholds which reliably tighten as we increase the order of the expansion.
The results for the first four orders are summarized in Tab.~\ref{tab:sphere_results} -- the first two areal orders are calculated analytically, for the rest we use a small scale Monte Carlo integration. 
\newline 
\begin{figure}[t]
	\includegraphics[width = 0.49\textwidth]{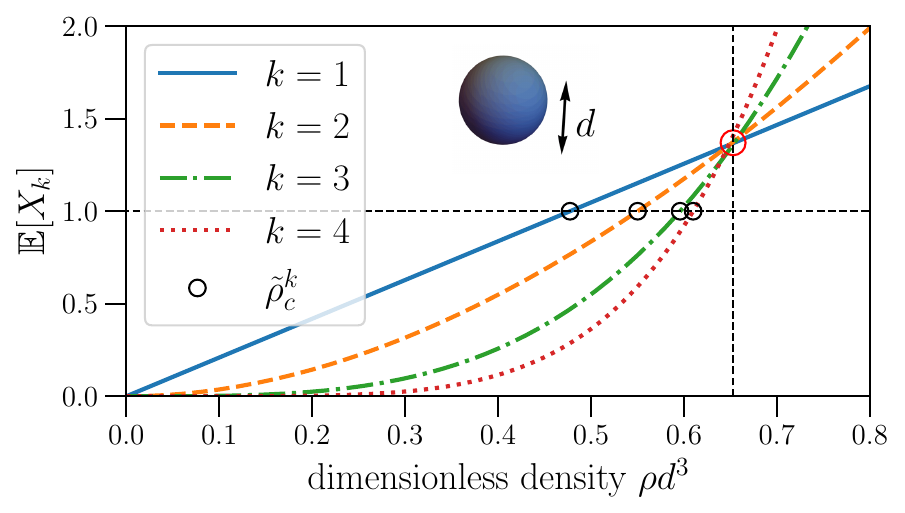}
	\caption{Mean surface activity $\expectation[X_k]$ of fully penetrable spheres as a branching process constrained to  a half-space. For $k \leq 2$ the $\mathbb{E}[X_k]$ can be computed analytically, for $k > 2$ we use small-scale simulations. 
		The circles mark the roots of $\expectation[X_k] = 1$ corresponding to the lower bounds listed in Tab.~\ref{tab:sphere_results}. The red circle contains all mutual intersections between the surface activities at different orders. The vertical dashed line demarcates the critical density determined with high-precision simulations \cite{lorenz2001precise}. }
	\label{fig:jwalking}
\end{figure}
However, despite a substantial improvement over the virial approximation, the lower bounds are still not particularly tight. This is because we have analysed branching from the origin rather than asymptotic growth.
The origin has a full $4 \pi$ solid angle available to branch to, whereas a particle on the surface of the growing cluster after $j$ branching iterations can only grow outward because the ball of radius $(j-1)d$ has already been specified.
If $j$ is large but finite we still find for any $\rho > 0$ an open particle $J$ on the surface of the cluster with finite probability. 
Thus, the system percolates if a branching process initiated at $J$ does not eventually terminate with probability 1. 
Yet, if $j$ is sufficiently large, the branching process starting at $J$ is effectively constrained to a half-space. 
By switching from the origin to $J$, we remove the vertex degree correlations caused by the artificial symmetry of the origin (like one additional branching direction on the homogeneous Bethe lattice in Fig.~\ref{fig:Bethe_lattices}),
getting rid of the explicit $k$-dependence $\xi_i^k$ in eq.~(\ref{eq:main}) by ``fast-forwarding to infinity''. 
Fig.~\ref{fig:jwalking} depicts the average surface activity for the constrained branching process. 
The resultant lower bounds to the percolation threshold $\tilde{\rho}_c^k$ have significantly tightened (see Tab.~\ref{tab:sphere_results}). 
Moreover, the mean surface activities for different orders $k$ mutually intersect each other in close proximity, at a density likewise extremely close to the literature critical density. 
This  pattern is familiar from finite size scaling analysis and exact for treelike networks (see e.g.~ref.~\cite{coupette2022exactly}). 

For general systems we expect the intersection point to drift slightly towards the percolation threshold.  
Nonetheless, equating the analytical expressions for orders $k=1$ and $k=2$ yields $\varrho_c \approx 0.6496$ agreeing with the literature value up to $0.5\%$. For a treelike system the different orders intersect at $\expectation[X_k] =  1$. Thus, the larger value of roughly $\expectation[X_k] \approx  1.37$ at the intersection reflects the inefficiency of the growth process due to loop formation indirectly quantifying the treelikeness of the continuum system. Generalizing the process to $D \geq 2$ dimensional embedding space, this value decreases with $D$ confirming the expected trend that percolation becomes more treelike in higher dimensions.

\subsection{Rodlike Particles}

\begin{figure}[t!]
	\includegraphics[width = 0.49\textwidth]{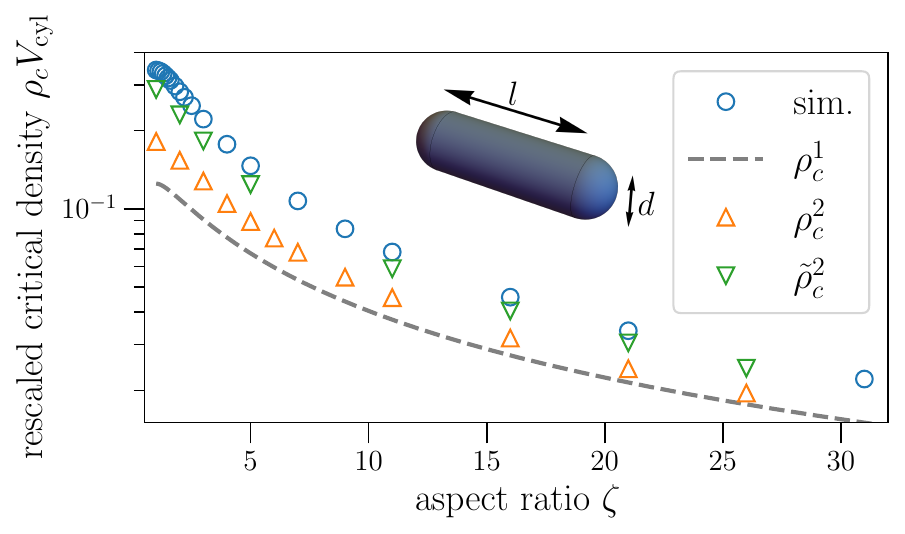}
	\caption{Percolation thresholds for penetrable spherocylinders with aspect ratio $\zeta$ and volume $V_{\mathrm{cyl}}$. Simulation results (blue circles) are taken from \cite{schilling2015percolation}. 
		The dashed line denotes the second virial approximation. 
		The areal approximants are calculated for linear chains of densely overlapping spheres ($V_{\mathrm{cyl}}$ is computed accordingly). }
	\label{fig:ar}
\end{figure}
Finally, we study the impact of particle anisotropy on the percolation threshold. 
Consider a penetrable spherocylinder of thickness $d$ and length $l+d$, $l$ being the length of the linear segment, and define its aspect ratio $\zeta := 1 + \frac{l}{d}$. 
In the Onsager limit $\zeta \rightarrow \infty$, the second virial approximation becomes exact \cite{kyrylyuk2008continuum}. This is not a particular strong statement as the critical density goes to 0 in this limit. Yet, for $\zeta < 100$ the approximation is inaccurate as simulation studies have shown \cite{berhan2007modeling,mutiso2012simulations,schilling2015percolation}. Heuristic corrections have been proposed by subjecting the critical number of nearest-neighbours to a power-law fit in the aspect ratio \cite{berhan2007modeling,mutiso2012simulations}. 
But the functional form of this fit is unjustified and does not agree with later simulation studies \cite{schilling2015percolation}. 
We can improve on the second virial approximation using the areal framework: we integrate out a ball of radius $\zeta$ and average the number of next-nearest neighbours yielding $X_2$. 
The solutions to $\mathbb{E}[X_2] = 1$ for different aspect ratios lead to the lower bounds $\rho_c^2$ illustrated in Fig.~\ref{fig:ar}. 
We observe only a slight improvement over the second virial approximation ($\rho_c^1$). 
Again, we can constrict the branching process to a half-space to find a massive tightening of the lower bound. 

The relative deviation to the simulation results is most pronounced for $\zeta = 2$ ($\lesssim 20\%$) and decreases monotonically with $\zeta$ ($\approx 10 \%$ at $\zeta = 21$). This is expected as the importance of loops in the formation of the percolating cluster diminishes with the aspect ratio ultimately approaching treelike topology in the Onsager limit.
This lowest order above second virial demonstrates that the severe shortcoming of the second virial approximation is not primarily the neglect of loops but rather the omission of vertex degree correlations. Indeed, the second virial approximation is inherently not as tight as it could be by eliminating the correlations induced by an overly symmetric choice of origin.
\newline
\newline
In summary, we introduced a general mapping of percolation problems onto branching processes. 
It is easily applicable to any type of percolation problem: pick an origin and define a notion of $k$-neighbourhood, construct the corresponding random variables $X_k$ and solve $\expectation[X_k] = 1$ to obtain rigorous lower bounds to the percolation threshold. 
If the origin is not representative for the asymptotic cluster growth, be it through additional symmetry, an external field or vertex correlations on a lattice, modify $X_k$ to reflect the asymptotic branching process. 
The resulting lower bounds to the percolation threshold include the first $k$ virial orders and extrapolate, in contrast to the Padé approximation, in a meaningful physical and controlled manner. 
Through consistency with the virial approximation we obtain the correct response of critical parameters to small parameter changes in combination with quantitatively accurate predictions. Our approach transfers principles of lattice percolation to the continuum, thus providing a unified framework for the prediction of critical parameters of connectivity percolation problems.

\begin{acknowledgments}
	We acknowledge funding by the German Research Foundation in the project 457534544. 
	Moreover, we thank Christoph Widder for constructive comments and helpful feedback on the manuscript.
\end{acknowledgments}

\section*{Data availability}

Results of the Monte Carlo integration for the calculation of higher order (areal) approximations (Figs.~\ref{fig:jwalking}, \ref{fig:ar} and 2 (SM) are available on reasonable request to Fabian Coupette. The remaining results are entirely analytical and readily reproduced with the equations provided in the article.

%\section*{Competing Interests}
%\noindent
%There are no competing interests to declare.
%
%\section*{Author contributions}
%\noindent
%F.C.: conceptualisation (equal), investigation \& writing (equal); T.S.: conceptualisation (equal), writing (equal), funding acquisition \& supervision.  

%\bibliography{literature}

\clearpage

\pagebreak

\makeatletter
\def\mysequence#1{\expandafter\@mysequence\csname c@#1\endcsname}
\def\@mysequence#1{%
	\ifcase#1\or 1\or 1\or 1\or 1\or 1\or 1\or 1\or 2\or 3\or 4\or 5\or 6\or 7\or 8\or 9\or 10\or 11\or 12\or 13\or 14\or 15\or 16\or 17\or 18\or 19\or 20\or 21\or 22\or 23\or 24\or 25\or 26\or 27\or 28\else\@ctrerr\fi}
\makeatother
\renewcommand\thesection{\mysequence{section}}

\setcounter{equation}{0}
\setcounter{table}{0}
\setcounter{page}{1}
\makeatletter
\renewcommand{\theequation}{S\arabic{equation}}
\renewcommand{\thefigure}{S\mysequence{figure}}
\renewcommand{\bibnumfmt}[1]{[S#1]}
\renewcommand{\citenumfont}[1]{S#1}
\renewcommand{\thetable}{S\arabic{table}}

\title{Supplemental Material for: A universally applicable approach to connectivity percolation
}

%\author{Fabian Coupette}
%\email{fabian.coupette@physik.uni-freiburg.de}
%\affiliation{Institute of Physics, University of Freiburg, Hermann-Herder-Stra{\ss}e 3, 79104 Freiburg, Germany}
%\author{Moritz B\"ultmann}
%\author{Andreas H\"artel}
%\affiliation{Institute of Physics, University of Freiburg, Hermann-Herder-Stra{\ss}e 3, 79104 Freiburg, Germany}
%\author{Tanja Schilling}
%\email{tanja.schilling@physik.uni-freiburg.de}
%\affiliation{Institute of Physics, University of Freiburg, Hermann-Herder-Stra{\ss}e 3, 79104 Freiburg, Germany}

\date{\today}

\begin{abstract}
	
	Part 1 discusses the problem of long range percolation on the integer line demonstrating how the framework developed in the main manuscript can be utilised to improve on the existing lower bounds to the percolation threshold.
	In part 2, we briefly outline the central elements of connectedness percolation theory and illustrate that the corresponding low-order virial approximations are inconclusive.
	
\end{abstract}

\maketitle

\onecolumngrid
\vspace{-1cm}
\section{Long range percolation on the integer line}\label{app:A}

\begin{figure}[h!]
	\includegraphics[width = 0.5 \textwidth]{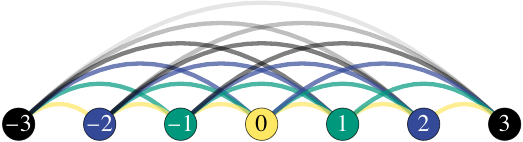}
	\caption{Long-range percolation on the integer line}
	\label{fig:1d_lr}
\end{figure}

Consider bond percolation on a complete graph with vertex set $\mathbb{Z}$. Each pair of vertices $(i,j)$ is connected by an edge which is open with a probability $g(i,j):=g(|i-j|)$ that depends on the distance between the adjacent vertices on the integer line $z := i-j$. The setup is illustrated in Fig.~\ref{fig:1d_lr}. In particular, we are interested in the family of power-law decaying edge probabilities
\begin{align}
	g(z) = \frac{p}{|z|^\sigma}
\end{align}
with $\sigma \in [1,2]$ and $p\in[0,1]$. For each $\sigma$ there is a critical $p_c(\sigma)$ corresponding to the first emergence of a percolating cluster, i.e., $p_c = \sup\{ p: \Theta(p)=0\}$. For $\sigma < 1$, the edge probability decays slow enough so that infinitely many open edges are expected to emanate from any site which means $p_c = 0$. Likewise, for $\sigma > 2$ there is almost surely a gap in the line that cannot be bridged unless $p = p_c = 1$.  For $\sigma \in [1,2]$, $p_c(\sigma)$ interpolates between the two trivial limits and the determination of this value makes for a nice illustration of the perspective to percolation we advocate in the main text.  

Since the shortest distance between any two vertices on the complete graph is 1, the notion of mutually exclusive vertex sets $V_k$ becomes meaningless. Instead, we define these sets for a given realization of the percolation problem. That means, $V_1$ is the set of vertices that are connected to the origin via one open edge. $V_2$ then comprises all the vertices that can be reached from the vertices in $V_1$ through one open edge and that have not been visited before. This breadth search protocol progressively generates mutually disjoint vertex sets $V_k$ and we can define $X_k = |V_k|$. Clearly, a terminating sequence $(X_k)_{k \in \mathbb{N}}$ corresponds to a finite cluster around the origin. We can cast the evolution of $X_k$ in the form of the generalized branching process
\begin{align}
	X_{k+1} = \sum^{X_{k}}_{i=1} \xi_i^k(\{\xi^k_j\}_{j<i}) \quad \text{with} \quad X_0 = 1 \; .
	\label{eq:mainS}
\end{align}  
The random variables $\xi_i^k$ derive from the distribution of open edges emanating from a vertex to vertices that have not been visited previously. Thus, every vertex that has been visited before reduces the expectation of $\xi_i^k$. Accordingly, all the random variables $(\xi_i^k)_{k \in \mathbb{N},1 \leq i \leq X_k }$ are anti-correlated. Neglecting correlation can hence only increase $X_k$.
\newline
\newline
Consider the branching process with all $\xi_i^k \equiv \xi$ where $\xi$ is distributed according to the distribution of open edges emanating from the origin. This branching process simply ignores that vertices have been visited before and overestimates $X_k$ compared to the original process for any vertex that is revisited. As this process is a true tree (a Galton-Watson branching process), it is critical if
\begin{align}
	\expectation[\xi] = 2 \sum_{i=1}^{\infty} g(i) = 2 \zeta(\sigma) = 1 \; ,
	\label{eq:lb}
\end{align}
featuring the Riemann $\zeta$-function.
\begin{figure}[b!]
	\includegraphics[width = 0.43 \textwidth]{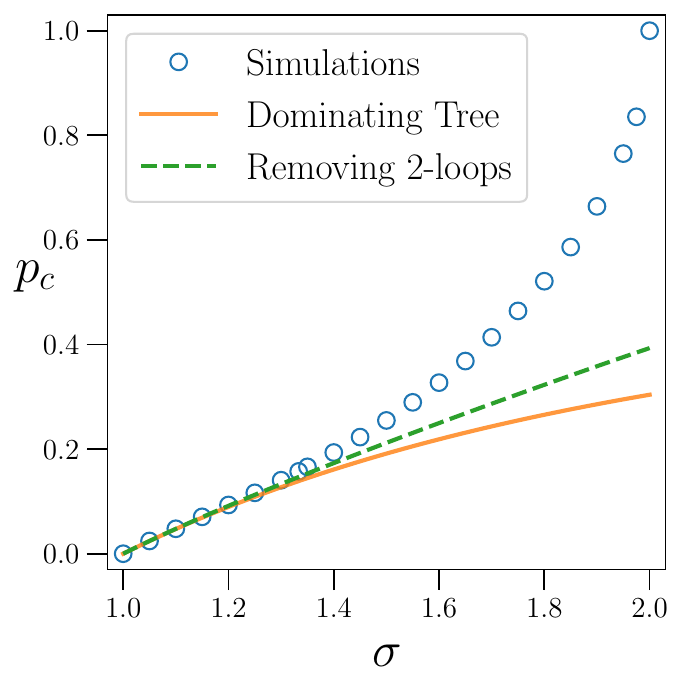}
	\includegraphics[width = 0.5 \textwidth]{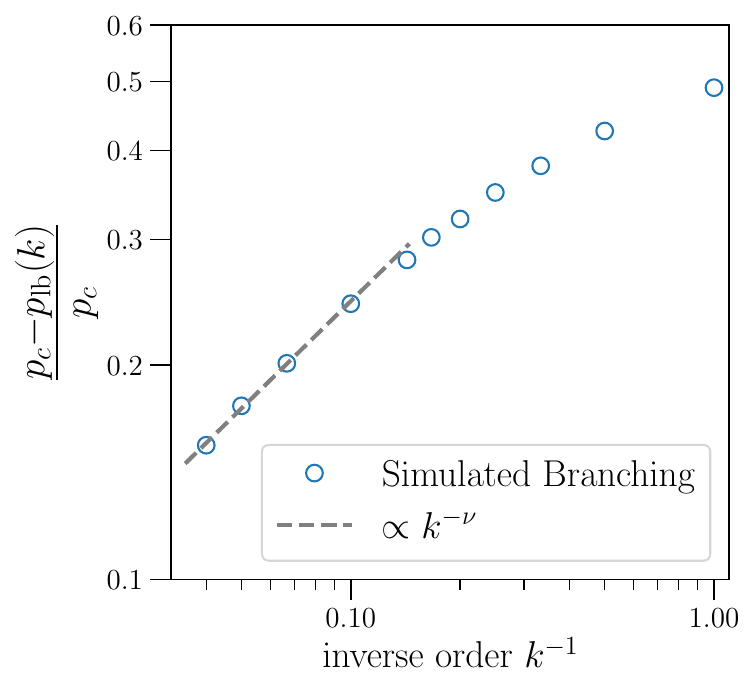}
	\caption{Left: Simulation results for long range percolation on the integer line compared to the analytic results of the two lowest order lower bounds for the percolation threshold generated by our framework. Right: Scaling of the tightness of the lower bound as a function of the order for $\sigma = 1.8$ in double logarithmic depiction. The lower bound scales with the correlation length critical exponent $\nu$.}
	\label{fig:lr_fs}
\end{figure}
This process dominates the generalized branching process we are actually interested in, i.e., the realizations of the percolation model form a subset of the trajectories that the uncorrelated branching process generates. That means, if our system is percolating the dominating branching process must be (super)critical as well and $p = \frac{1}{2 \zeta(\sigma)}$ must be a lower bound of the percolation threshold. This bound has been determined before  \cite{schulman1983longS} by very similar arguments and is displayed in Fig.~\ref{fig:lr_fs} in comparison with simulation results of that model reported in \cite{gori2017oneS}. Yet, our approach now allows us to systematically improve on that lower bound by eliminating trajectories from the dominating tree that are precluded in the evolution of the actual surface activity. For example, we can demand that an open vertex  $v \in V_k$ cannot branch back immediately to its parent vertex (the vertex in $V_{k-1}$ corresponding to the random variable that generated $v$ as offspring). For a growing cluster this branching step would form a 2-loop that artificially inflates the surface activity in the dominating tree. The random variable $\xi$ that corresponds to this refined branching process simply excludes the origin as branching destination for all the vertices $V_1$. This translates after some algebra \cite{coupette2023percolationS} to the improved criticality criterion   
\begin{align}
	&\expectation[X_2 | X_1 = 1] =
	2 \sum_{z=1}^{\infty} g(|z|) - \nonumber\\ 
	&\left( \sum_{z \in \mathbb{Z} \setminus \{ 0 \}  } \frac{g(|z|)}{1-g(|z|)} \right)^{-1} \sum_{z \in \mathbb{Z} \setminus \{ 0 \}  } \frac{g(|z|)^2}{1-g(|z|)} = 1 \; ,
	\label{eq:long_range_first _order}
\end{align}
which yields the dashed line in Fig.~\ref{fig:lr_fs}. As consistency requires, the refined bound is strictly in between the simulation results and the lower bound of eq.~(\ref{eq:lb}). The improvement is not particularly impressive but in view of the fact that we only remove one of infinitely many branching destinations for any vertex in $V_k$, it still is remarkable. Moreover, we only removed the smallest loops from the dominating tree. Yet, the next orders of the approximation scheme already requires sums over power sets of the integers rendering the analytic pursuit rather tedious. However, we may still use brute force simulations to evaluate the hierarchy of lower bounds corresponding to 
\begin{align}
	\expectation[X_k] = 1
\end{align}
with $X_k$ evolving to the actual generalized branching process of eq.~(\ref{eq:mainS}). The tightness of the corresponding lower bounds as a function of $k$ are depicted for $\sigma = 1.8$ in the right panel of Fig.~\ref{fig:lr_fs}. The results illustrate that the lower bound scales effectively like the correlation length with $\nu = \frac{1}{2}$ being the corresponding mean field critical exponent. 

\section{Virial Expansion for Percolation}
\label{app:B}
\begin{table*}[t]
	\centering
	\caption{$C_i^\dagger(\bs{0})$ resulting from the individual diagrams of eq.~(\ref{eq:c_dagger_expansion}) in the same order. The last expression utilizes a result from \cite{nijboer1952radialS}.
	} 
	\label{tab:sphere_results2}
	\begin{ruledtabular}
		\begin{tabular}{c l l} \toprule
			virial order &  diagram index & $C^\dagger(\bs{0})$  \\\midrule
			2 & I  & $\frac{4}{3} \pi d^3$ \\
			3 & II & $\frac{5}{6} \pi^2 \rho d^6 $ \\
			4 & III & $\frac{2176}{2835} \pi^3 \rho^2 d^9$ \\
			4 & (IV - VII) & $\frac{2357}{11340} \pi^3 \rho^2 d^9$    \\
			4 & V \& VI &  $\frac{6347}{11340} \pi^3 \rho^2 d^9$   \\
			4 & (IX - VIII) & $\frac{\pi^2 \rho^2 d^9}{22680}\left[4903 \pi +18(292 \sqrt{2}-3126 \arccot(\sqrt{2})+681 \arccsc(3) + 1248 \arctan(\frac{5}{\sqrt{2}}) )\right]$   \\\bottomrule
		\end{tabular}
	\end{ruledtabular}
\end{table*}
\begin{figure}[h]
	\includegraphics[width = 0.49\textwidth]{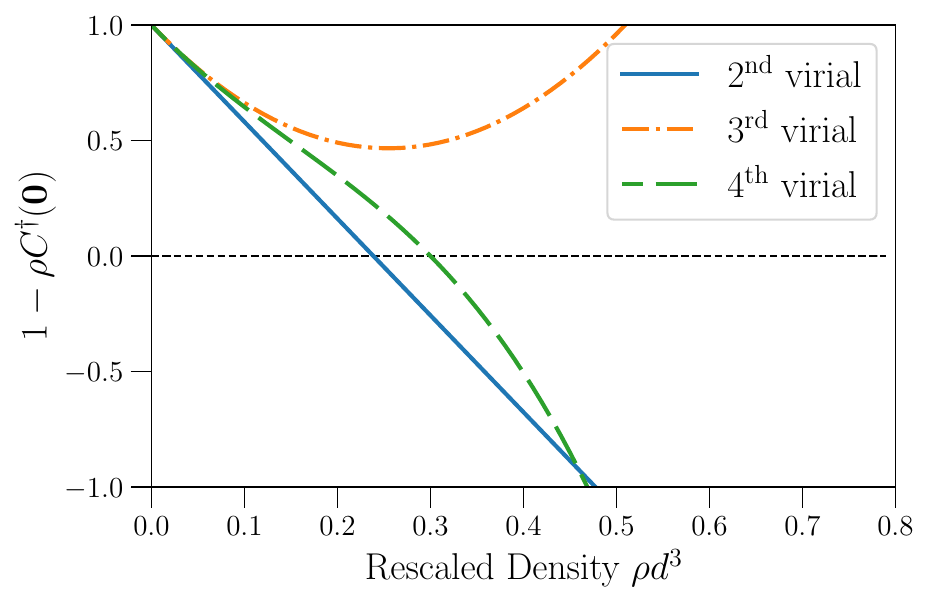}
	\caption{Virial approximations for percolation of fully penetrable spheres}
	\label{fig:virials}
\end{figure}
Connectedness percolation theory employs the cluster integral representation of the pair-correlation function in liquid state theory and adds the information of connectedness microscopically by modifying the Mayer $f$ bond. The pair-connectedness is the analogue of the pair-distribution function in that the expression
\begin{align}
	P(\bs{r},\bs{r}') \dd \bs{r} \dd \bs{r}'
\end{align}
describes the probability to simultaneously find particles in respective volume elements $\dd \bs{r}$ and $\dd \bs{r}'$ which are also part of the same connected component.
Accordingly, the mean cluster size is given by
\begin{align}
	S(\bs{r}) = 1 +  \int \dd \bs{r}' \rho(\bs{r'}) P(\bs{r},\bs{r}') \; ,
\end{align}
which simplifies for a homogeneous system to
\begin{align}
	S= 1 + 4 \pi \rho \int \dd r \, r^2  P(r) = 1 + \rho \hat{P}(\bs{0}) \; ,
	\label{eq:MCS}
\end{align}
with $r = |\bs{r}'-\bs{r}|$ and $\hat{P}(\bs{k})$ denoting the Fourier transform of $P(\bs{r})$.
The connectedness Ornstein-Zernike equation relates the pair connectedness $P$ to the direct connectedness $C^\dagger$ 
\begin{align}
	P(\bs{r},\bs{r}') = C^{\dagger} (\bs{r},\bs{r}')  + \int \dd \bs{r}'' \, C^{\dagger} (\bs{r},\bs{r}'')  \rho^{(1)}(\bs{r}'') P(\bs{r}'',\bs{r}') \; .
	\label{eq:cOZ}
\end{align}
For a homogeneous system eq.~(\ref{eq:cOZ}) is reduced to convolution which results in an algebraic equation in Fourier space
\begin{align}
	\hat{P} = \hat{C}^\dagger + \rho \, \hat{C}^\dagger \hat{P} \; . 
\end{align} 
This equation implies
\begin{align}
	\hat{P}(\bs{0}) = \frac{\hat{C}^\dagger(\bs{0})}{1 - \rho \, \hat{C}^\dagger(\bs{0}) } \; ,
	\label{eq:cozF}
\end{align}
so that, referring to eq.~(\ref{eq:MCS}), the mean cluster size diverges if the right hand side of eq.~(\ref{eq:cozF}) becomes singular. Assuming $C^\dagger(\bs{0})$ is finite, the smallest positive real root of 
\begin{align}
	1 - \rho \; C^\dagger(\bs{0}) = 0
	\label{eq:root}
\end{align}
corresponds to the percolation threshold. The virial expansion of $C^\dagger$ in $f^\dagger(\bs{r},\bs{r}') = \Theta(d - |\bs{r}-\bs{r}'|)$ bonds and $\rho$ circles up to fourth virial order is given by

\begin{align}
	\begin{tikzpicture}[scale=0.6]
		\node at (-2.5,0) {$C^\dagger(\bs{r},\bs{r}')$};
		\node at (-1.15,0) {$=$};
		\node at (0,0) {\includegraphics[scale=0.12]{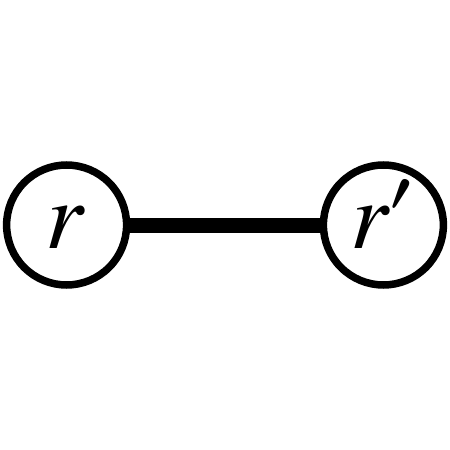}};
		\node at (1,0) {$-$};
		\node at (1.5,0) {$1$};
		\node at (2.5,0) {\includegraphics[scale=0.12]{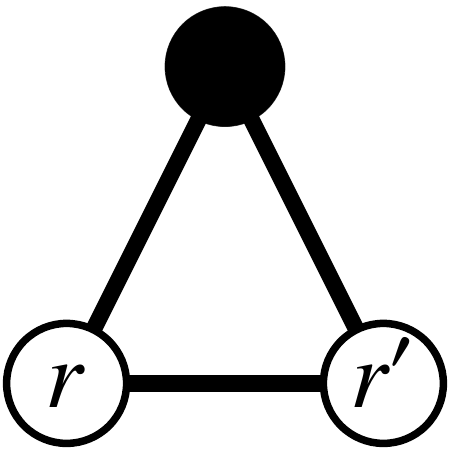}};
		\node at (3.5,0) {$-$};
		\node at (4,0) {$1$};
		\node at (5,0) {\includegraphics[scale=0.12]{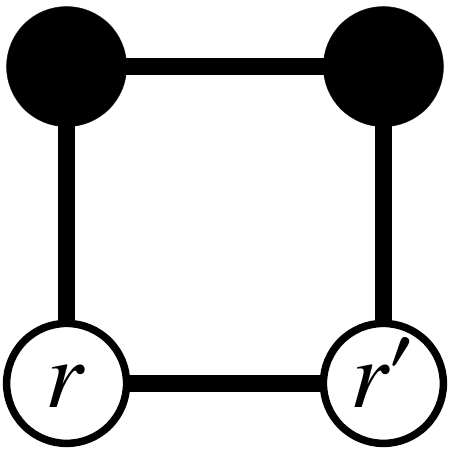}};
		\node at (6,0) {$-$};
		\node at (6.5,0) {$1$};
		\node at (7.5,0) {\includegraphics[scale=0.12]{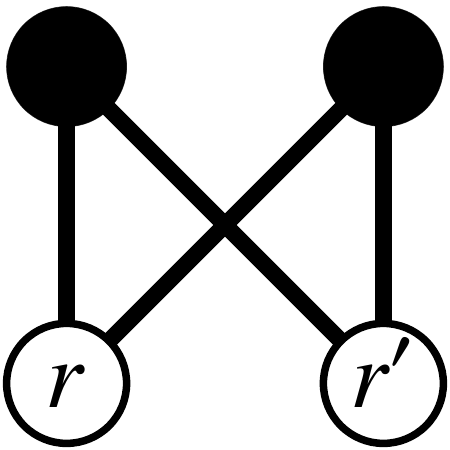}};
		\node at (8.5,0) {$+$};
		\node at (9,0) {$1$};
		\node at (10,0) {\includegraphics[scale=0.12]{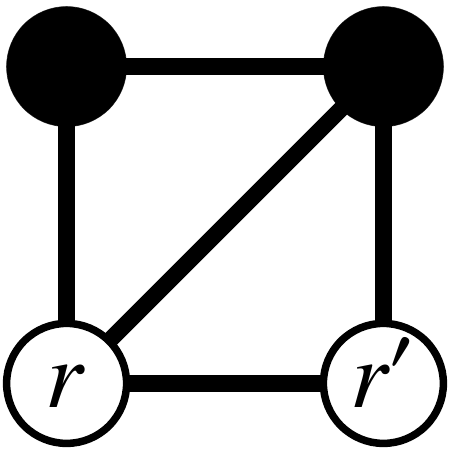}};
		\node at (-1.5,-2) {$+$};
		\node at (-1,-2) {$1$};
		\node at (0,-2) {\includegraphics[scale=0.12]{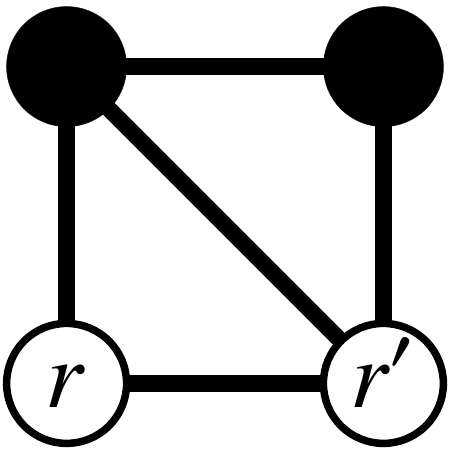}};
		\node at (1,-2) {$+$};
		\node at (1.5,-2) {$1$};
		\node at (2.5,-2) {\includegraphics[scale=0.12]{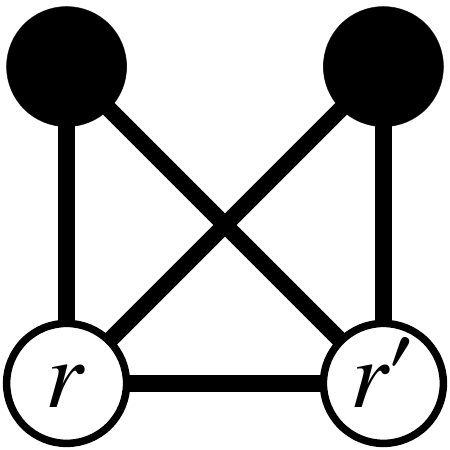}};
		\node at (3.5,-2) {$+$};
		\node at (4.0,-2) {$2$}; 
		\node at (5.0,-2) {\includegraphics[scale=0.12]{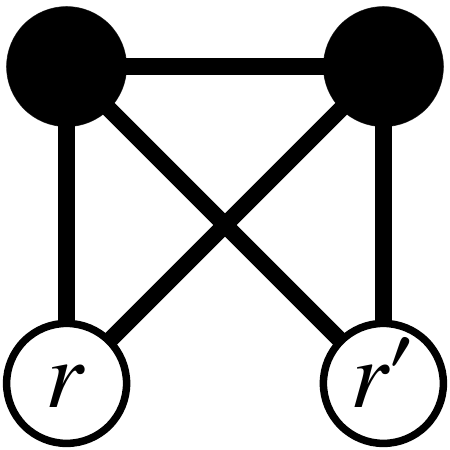}};
		\node at (6.0,-2) {$-$};
		\node at (6.5,-2) {$2$}; 
		\node at (7.5,-2) {\includegraphics[scale=0.12]{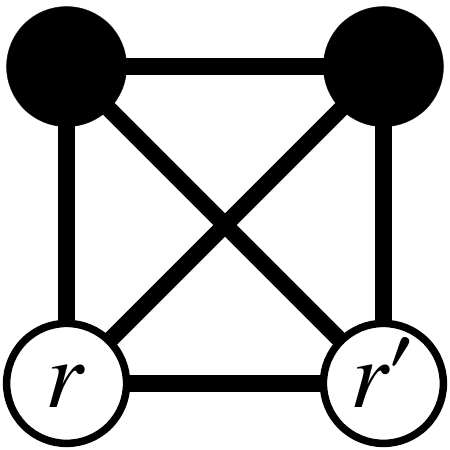}};
		\node at (8.5,-2) {$+$};
		\node at (10.0,-2) {$\mathcal{O}(\rho^3)$};
	\end{tikzpicture}
	\; ,
	\label{eq:c_dagger_expansion}
\end{align}

The constituent diagrams have been calculated previously for the virial expansion of the structure of a hard sphere fluid for which the Mayer f bond is given by $f_{\mathrm{HS}} \equiv -f^{\dagger}$. Yet, as commonly only the complete virial coefficients are reported, we list the analytic forms of the required integrals in table \ref{tab:sphere_results2}. The corresponding results for the left hand side of eq.~(\ref{eq:root}) are illustrated in Fig.~\ref{fig:virials}.  
The third virial approximation does not yield a real root and hence does not predict a percolation threshold at all. The large discrepancy between all orders for rescaled densities exceeding $\rho d^3 \approx 0.2$ underline that higher order terms have substantial impact and the series cannot be reliably truncated at fourth (and presumably not even at much higher) orders. Importantly, a Padé approximation does not resolve the issue that we have not included any of the relevant diagrams at densities close to the actual percolation threshold $\rho_c d^3 \approx 0.6530$. It only extrapolates in a more convenient way without any additional insight.

\end{document}